\documentclass[12pt,preprint]{aastex}

\def\lta{\,\raise 0.3 ex\hbox{$ < $}\kern -0.75 em
 \lower 0.7 ex\hbox{$\sim$}\,}
\def\gta{\,\raise 0.3 ex\hbox{$ > $}\kern -0.75 em
 \lower 0.7 ex\hbox{$\sim$}\,} 
\newcommand{\be}{\begin{equation}}
\newcommand{\ee}{\end{equation}}

\begin{document}

\title{Core-Accretion Model Predicts Few Jovian-Mass Planets Orbiting Red Dwarfs}   

\author{Gregory Laughlin$^1$, Peter Bodenheimer$^1$, and Fred C. Adams$^{2,3}$}

\affil{$^1$Lick Observatory, University of California, Santa Cruz, CA 95064 \\ 
laughlin@ucolick.org, peter@ucolick.org}

\affil{$^2$Michigan Center for Theoretical Physics, Physics Department \\ 
University of Michigan, Ann Arbor, MI 48109, fca@umich.edu}  

\affil{$^3$Astronomy Department, University of Michigan, Ann Arbor, MI 48109} 

\begin{abstract} 

The favored theoretical explanation for giant planet formation -- 
in both our solar system and others -- is the core accretion model
(although it still has some serious difficulties). In this scenario,
planetesimals accumulate to build up planetary cores, which then
accrete nebular gas. With current opacity estimates for protoplanetary
envelopes, this model predicts the formation of Jupiter-mass planets
in 2--3 Myr at 5 AU around solar-mass stars, provided that the surface
density of solids is enhanced over that of the minimum-mass solar
nebula (by a factor of a few).  Working within the core-accretion
paradigm, this paper presents theoretical calculations which show 
that the formation of Jupiter-mass planets orbiting M dwarf stars is 
seriously inhibited at all radial locations (in sharp contrast to
solar-type stars). Planet detection programs sensitive to companions
of M dwarfs will test this prediction in the near future.

\end{abstract} 


\section{Introduction} 

Over 100 planets have been detected orbiting Sun-like stars and the
distributions of extrasolar planetary masses and orbital properties
are remarkably diverse (Marcy et al. 2003; Udry, Mayor, \& Queloz
2003).  The leading mechanism for explaining the origin of both the
extrasolar planets and our solar system's Jovian planets is the
core-accretion process, in which collisional accumulation of icy
planetesimals builds planetary cores with mass $M \sim 5-15 \,
M_{\oplus}$. These cores then accrete nebular gas and reach Jovian
mass.  A long-standing difficulty with the core-accretion hypothesis
was that the estimated time required for the core to accrete $1~
M_{\rm JUP}$ of gas exceeded the observed circumstellar disk lifetimes
(Pollack et al. 1996). This issue motivated discussion of
gravitational instability as an alternative formation process (Boss
2000).  However, updated estimates of the opacity in protoplanetary
envelopes imply that Jupiter-mass planets readily form via
core-accretion in solar-metallicity disks within 2-3 Myr at radii $a
\sim 5 {\rm AU}$ around solar-mass stars (Hubickyj, Bodenheimer, \&
Lissauer 2003). This relatively short formation time results if the 
surface density of solid material in the disk is assumed to be three
times greater than that of the minimum-mass solar nebula (MMSN). The
resulting core mass was about 16 $M_\oplus$, somewhat larger than the
maximum deduced core mass (10 $M_\oplus$) for Jupiter (Wuchterl,
Guillot \& Lissauer 2000). To explain the low core mass of Jupiter, a
cutoff of solid accretion beyond a certain core mass is required, and
can be explained by nearby planetary embryos that compete for
available solid material. An independent calculation with similar
opacities (Inaba \& Ikoma 2003; Inaba, Wetherill, \& Ikoma 2003) shows
that in a disk with eight times the solid surface density of the MMSN,
a core of 25 $M_\oplus$ can form in 1 Myr at 5 AU, so that the total
formation time is 2--3 Myr. In this model, the fragmentation of
planetesimals and the enhancement of the solid accretion rate due to
the gaseous envelope (primarily from gas drag) are taken into account,
although the main gas accretion phase is not calculated. If the solid
surface density is reduced to four times that of the MMSN, an 8
$M_\oplus$ core can still form in 5 Myr. These results imply that
somewhat special circumstances are required for the core accretion
model to explain the properties of Jupiter. Assuming that the core
accretion model can explain the formation of Jovian planets around
solar-mass stars, this paper addresses the question of whether or not
Jovian planets can form within disks that orbit around M-dwarfs (low
mass stars with $M_\star \la 0.4 M_\odot$).

Observational surveys are shifting our view of extrasolar planets from
an anecdotal collection of individual systems (e.g., 51 Peg, $\upsilon$ 
And, or 47 UMa) to a fuller statistical census in which categories and
populations of planets can be clearly delineated (e.g., Marcy \&
Butler 1998; Marcy, Cochran, \& Mayor 2000; Udry et al. 2003; Marcy et
al. 2003).  This emerging statistical view is vital for improving our
understanding of the planet formation process, and to see how our own
solar system fits into the galactic planetary census.  
One of the most remarkable statistical results to emerge from planet
searches is that stars with observed extrasolar planets tend to have
metallicities that are more than twice that of the average Population
I star in the immediate galactic neighborhood (Fischer \& Valenti
2003; Butler et al. 2000).  On the other hand, low metallicity stars
are deficient in currently detectable ($P<8$ yr) Jovian-mass planets
(Sozzetti et al. 2004). This connection between planets and host-star
metallicity can be interpreted as evidence in favor of the core
accretion hypothesis, although Sigurdsson et al. (2003) argue that it
can also be interpreted as evidence in favor of migration.  Metal-rich
circumstellar disks have a higher surface density of solids and cores
can readily reach the $5-15 M_{\oplus}$ threshold required for rapid
gas accretion. This letter shows that the core-accretion process makes
a similar -- and readily testable -- prediction of the relative
frequency of Jovian-mass planets as a function of stellar mass.
Compared to disks around solar-mass stars, the circumstellar disks
orbiting red dwarfs ($M_\star < 0.5 M_{\odot}$) should be less
efficient in producing Jupiter-mass planets.

\section{Theoretical Model of Planet Formation} 

This theoretical treatment assumes that planets form within a
circumstellar disk with the following properties. The surface density
$\sigma(r)$ = $\sigma_{in}(r_{in}/r)^{3/2}$, where $r_{in}$ and $r_d$
are the inner and outer disk radii, and $\sigma_{in}=(M_d(t)/4\pi
r_{in}^2) [(r_d/r_{in})^{1/2}-1]^{-1}$ is the normalization factor
required to obtain a total disk mass $M_d(t)$. We specify the time
dependence through a disk depletion factor $f_\sigma(t)=1/(1+t/t_0)$
so that $M_d(t)=M_d(0){f}_{\sigma}(t)$.  Observations of protostellar
disks (e.g., Briceno et al. 2001) suggest that $t_0=10^{5}$ yr and
$M_{d}(0)=0.05\,M_\star$ are reasonable fiducial values. The
temperature distributions for both viscously evolving accretion disks
and flat, passively irradiated disks have nearly the same power-law
form, $T_{d}(r)=T_{d \star}(R_{\star}/r)^{3/4}$, where $T_{d \star}$
is related to the stellar surface temperature by a geometrical factor
($T_{d \star}/T_\star \approx [2/3 \pi]^{1/4}$ for a flat disk -- see
Adams \& Shu 1986). This model uses disks that are flat and passive, 
and assumes that the disk is isothermal in the vertical direction. 
The effective temperature $T_\star$ of the star is related to the
stellar radius $R_\star$ and luminosity $L_\star (t,M_\star)$ through
$T_\star (t,M_\star)= [L_\star (t,M_\star)/4\pi R_\star^{2} \sigma]^{1/4}$. 
We adopt $T_\star(t,M_\star)$ and $L_\star(t,M_\star)$ from published
pre-main-sequence stellar evolution tracks (D'Antona \& Mazzitelli 1994). 

We use a Henyey-type code (Henyey, Forbes, \& Gould 1964) to compute
the contraction and buildup of gaseous envelopes surrounding growing
protoplanetary cores embedded in our evolving model disk. Our method
(Kornet, Bodenheimer, \& R\'o\.zyczka 2002; Pollack et al. 1996)
adopts recent models for envelope opacity (Podolak 2003) which
includes grain settling (see also Hubickyj et al. 2003). The
calculation is simplified (compare with Pollack et al. 1996) in that
it uses a core accretion rate of the form $dM_{\rm core}/ dt = C_1
\pi \sigma_s R_c R_h \Omega$ (Papaloizou \& Terquem 1999), where
$\sigma_s$ is the surface density of solid material in the disk,
$\Omega$ is the orbital frequency, $R_c$ is the effective capture
radius of the protoplanet for accretion of solid particles, $R_h = a
[M_p/(3M_{\star})]^{1/3}$ is the tidal radius of the protoplanet, and
$C_1$ is a constant near unity. An important feature of our present
model is that the outer boundary conditions for the planet include the
decrease in the background nebular density and temperature with time. 

The results of this planet formation calculation are illustrated in
Figure 1. The first calculation is for a disk around a 1 $M_\odot$
star with an initial solid surface density $\sigma_s = 11.5$ g
cm$^{-2}$ at $a = 5$ AU, about four times that of the MMSN. This value
is based on an initial gas-to-solid ratio of 70 in the disk; later on,
$\sigma_s$ decreases with time as mass accretes onto the growing
planet. A Jupiter-mass planet forms in 3.25 Myr with a core mass of 18
$M_\oplus$, consistent with the results of Hubickyj et al. (2003). The
second calculation is for a disk around an $M_\ast$ = 0.4 $M_\odot$
red dwarf with an initial solid surface density of 4.5 g cm$^{-2}$ at
$a = 5$ AU. Formation of a Jupiter-mass planet does not occur: Planet
growth has reached only $M_P = 14 M_\oplus$ at $t$ = 10 Myr. At later
times, much less than 1 $M_{JUP}$ remains in the entire disk. The
resulting planet is similar in mass, size, and composition to Uranus
and Neptune.

Figure 2 shows that Jovian planet formation around a $0.4 M_{\odot}$
star is also thwarted at radii of 1 AU and 10 AU. The lower surface
densities ($\sigma_s$) and longer orbital timescales ($\Omega =
\sqrt{GM_{\star}/a^3}$) of M-star disks more than offset the increase
in tidal radius $R_h$ and lead to a greatly reduced capacity for
forming Jovian-mass planets within the core-accretion paradigm. Also,
the reduced core mass found in the M-star disk results in much longer
times for the accretion of the gaseous envelope (Pollack et al. 1996).
Figure 2 indicates that although M stars (red dwarfs) have a limited
ability to form Jupiter-mass planets, the formation of Neptune-like
objects and terrestrial-type planets should be common around these
low-mass stars. This finding is the main result of this paper.
Furthermore, the final sizes of these objects correlates with the
surface density of solids (dust and ice) in the precursor
protoplanetary disk and should thus depend on the host star
metallicity.

\section{Potential Tests of the Theory} 

A number of detection methods can potentially confirm a paucity of
Jupiter-mass planets orbiting M-stars, and also detect Neptune-mass
objects. Given an adequate time baseline, the Doppler radial velocity
method (Marcy \& Butler 1998) will readily determine whether
Jupiter-mass planets are common around red dwarfs. A Neptune-mass
planet in a circular, $i=90^{\circ}$, $a$=3 AU orbit around a $0.4
M_{\odot}$ primary would have a period $P=8.2 \, {\rm yr}$, and would
induce a stellar radial velocity half-amplitude $K=1.6\, {\rm
m\,s^{-1}}$. Such an object would be marginally detectable using
current RV precision of $3\,{\rm m\,s^{-1}}$, assuming a high sampling
cadence. The California and Carnegie Planet Search radial velocity
program has detected one planetary system orbiting the M-dwarf star GJ
876 and $\sim100$ other red dwarfs are currently under surveillance. 

The GJ876 system (Marcy et al. 2001), with $M_{\star}$ $\sim0.3$
$M_{\odot}$, contains two Jovian planets ($M_c\sin i=0.6M_{\rm JUP}$,
$M_b \sin i=1.9M_{\rm JUP}$) with periods $P_c\sim30\,{\rm d}$ and
$P_b\sim60\,{\rm d}$. The system displays clear signs of having
undergone migration (Lee \& Peale 2002), indicating that abundant gas
was present in the system when the planets formed. We argue that
standard core-accretion theory predicts that systems such as GJ 876
are drawn from the extreme high-mass end of the circumstellar disk
mass distribution, and will thus be intrinsically rare. In the
alternative formation mechanism (gravitational instabilities; see Boss
2000), the growth rate depends primarily on the ratio of disk mass to
star mass and instabilities need not be suppressed in disks
surrounding low mass stars (compared to disks around solar-type
stars). Systems like GJ 876 may turn out to be examples of giant
planet formation via gravitational instability.  On the other hand,
gravitational fragmentation will not generally produce ice giants
(such as Neptune), so the discovery of ice giants in extrasolar
systems would be an important confirmation of the core-accretion
hypothesis.

Microlensing is another promising method for determining the census of
both high and low mass planets orbiting M dwarfs. Recent work (Bond et
al. 2004) reports an unusual light-curve for a G-type source star in
the direction of the galactic bulge. The observed light curve is
consistent with the passage of an optically faint binary lens with
mass ratio $\mu=0.0039^{+11}_{-07}$ for the lensing system. Stellar
population models of the galactic disk (Han \& Gould 1996) indicate a
$\sim90$\% a-priori probability that the optically faint lens primary
is an M-dwarf (in which case the planet mass is of order $M_P \sim 1.5
M_{\rm JUP}$) and a $\sim10$\% probability that the primary is a white
dwarf (implying a somewhat higher planet mass). In either case, the
projected sky separation of the primary--planet system is $\sim3$ AU.
Our calculations imply that M-dwarfs should rarely harbor Jupiter-mass
planets and favor the possibility that the lensing system has a white
dwarf primary. This prediction can be tested within $\sim10$ years as
proper motion separates the source star from the lens. We also predict
that the mass spectrum of microlensed planets (drawn largely from
M-dwarf primaries) should be shifted dramatically toward Neptune-mass
objects compared to the mass spectrum of planets found by the ongoing
RV and Transit surveys (which draw predominantly from primaries of
roughly solar-mass).

Transits provide another method for detecting planets. A Neptune-mass
object in central transit around a 0.3 $M_{\odot}$ M dwarf produces a
photometric dip of approximately 1.5\%. Such events are easily
observed from the ground using telescopes of modest aperture (Henry et
al. 2000, Seagroves et al. 2003). The forthcoming Kepler mission (Koch
et al. 1998) will monitor $\sim$3600 M dwarf stars with $m_{V}<16$
over its 4-year lifetime, and will easily detect transits of objects
of Earth size or larger in orbit around M dwarf stars. Assuming that
icy core masses of $M \sim 1 M_\oplus$ can accrete at $a \sim 1$ AU,
the Kepler sample size will be large enough to provide a statistical
test of our hypothesis.

\section{Conclusion} 

Within the context of the core-accretion process, this paper argues
that M dwarfs face a number of difficulties in producing Jupiter-mass
planets. Our principal conclusion is that Jovian planets should be
rare in solar systems orbiting red dwarfs, but Neptune-like objects
and terrestrial planets should be common around these low-mass stars
(see Figures 1 and 2). Since disks orbiting solar-type stars readily
produce giant planets (Figure 1), we predict that planet properties
should vary with the mass of the central star. This result is
straightforward to understand: Giant planet formation must take place
before the disk gas is dispersed, and the dynamical time scales in M
dwarf systems are longer (the Keplerian orbit time scales like
$[M_\star/M_\odot]^{-1/2}$).  Giant planets form in the outer solar
system where ices are frozen onto the rocky building blocks, but the
disks around M dwarfs have a much lower surface density in the realm
of the nebula beyond the snowline. The mass supply is smaller by a
factor of $M_\star/M_{\odot}$, the growth rate for planetesimal
formation is smaller by a factor $(M_\star/M_{\odot})^2$, and the late
stages of planetesimal accumulation proceed $M_\star/M_{\odot}$ times
slower. All of these effects tend to impede giant planet formation,
which must take place before the gas in the nebula is removed.

In addition to the effects calculated herein, planets forming around
red dwarfs face additional hurdles. Most stars form in groups and
clusters, where external radiation from other nearby stars can
efficiently drive mass loss from disks around M stars (Adams et al.
2004). In their youth, M stars are almost as bright as solar type
stars, but their gravitational potential wells are less deep; this
combination of properties allows the inner disks to be more readily
evaporated. For M stars, compared with solar-type stars,
photoevaporation due to both external radiation fields and radiation
from the central star can be more effective by a factor of 10 -- 100
(depending on the environment), so the gas in the nebula can be much 
shorter lived. Star forming regions are dynamically disruptive due to
passing binary stars and background tidal forces; these influences
affect circumstellar disks and planetary systems around M stars more
effectively than in systems anchored by larger primaries.  Finally,
all of these difficulties affect not only planet formation, but also
planet migration, indicating that very short-period Jovian-mass
planets should be especially rare near M-stars.

\newpage  

It is a pleasure to thank Geoff Marcy, Paul Butler, Debra Fischer, and
Steve Vogt for useful conversations. We would also like to thank Alan 
Boss -- the referee -- for his prompt and insightful report. This work 
was supported at UCSC by NASA through the Terrestrial Planet Finder
Precursor Science Program (grant NNG04G191G to GL) and through the
Origins of Solar Systems Program (grant NAG5-13285).  This work was
supported at the Univ. Michigan (FCA) by the Michigan Center for
Theoretical Physics and by NASA through the Terrestrial Planet Finder
Mission and the Astrophysics Theory Program.

\begin{figure}
\plotone{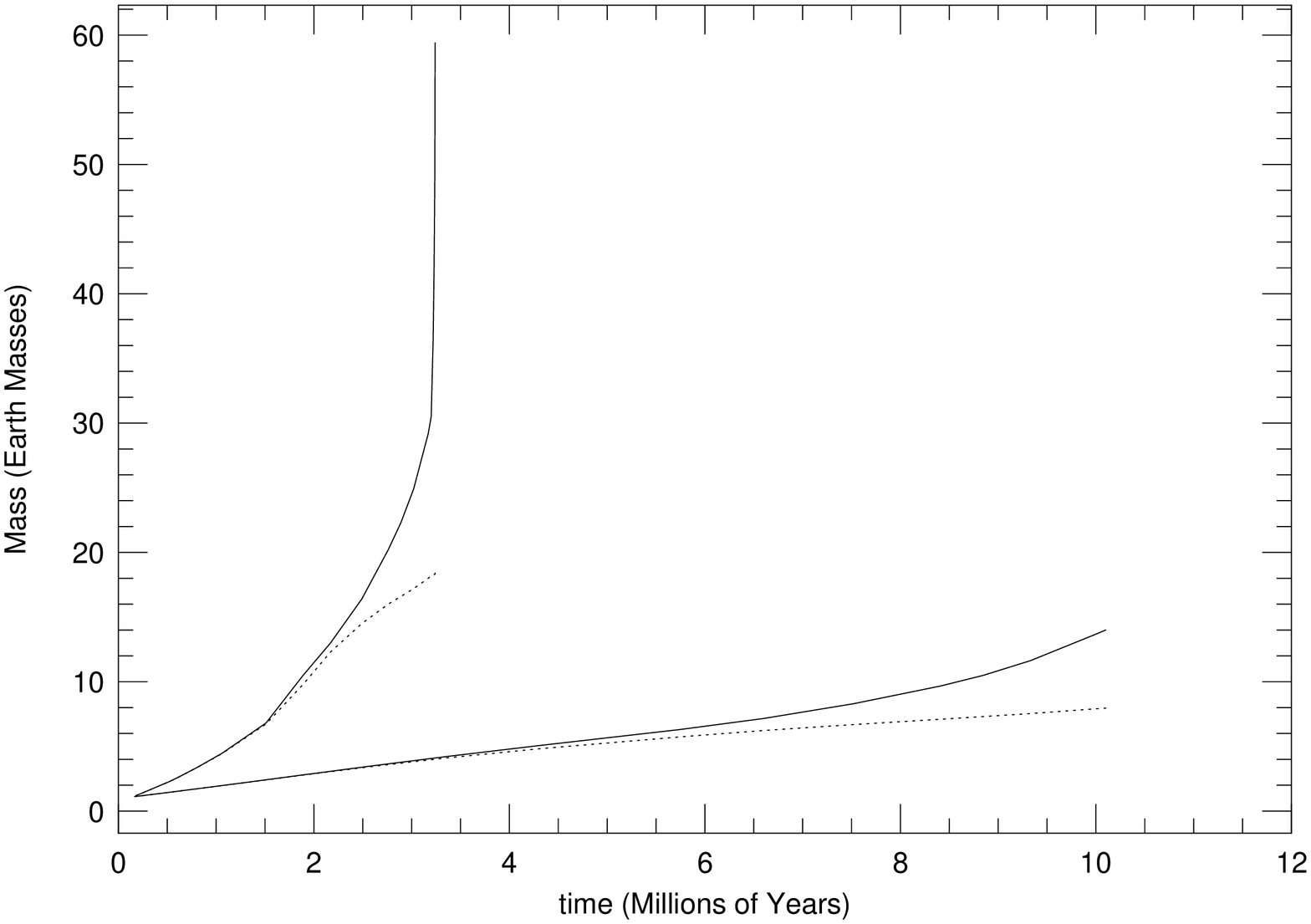}
\caption{Growth of the core and envelopes of planets at 5.2 AU in disks 
orbiting stars of two different masses.  The upper curves show the
time-dependent core mass (dotted curve) and total mass (solid curve)
for a planet forming in a disk surrounding a $1 M_{\odot}$ star. The
lower curves show the time dependence of the core mass (dotted curve)
and total mass (solid curve) for a planet forming in a disk around a 
$0.4 M_{\odot}$ star. After 10 Myr, the disk masses become extremely 
low, which effectively halts further planetary growth. The planet 
orbiting the M star gains its mass more slowly and stops its growth 
at a relatively low mass $M \approx 14 M_{\oplus}$. } 
\end{figure}
\clearpage

\begin{figure}
\plotone{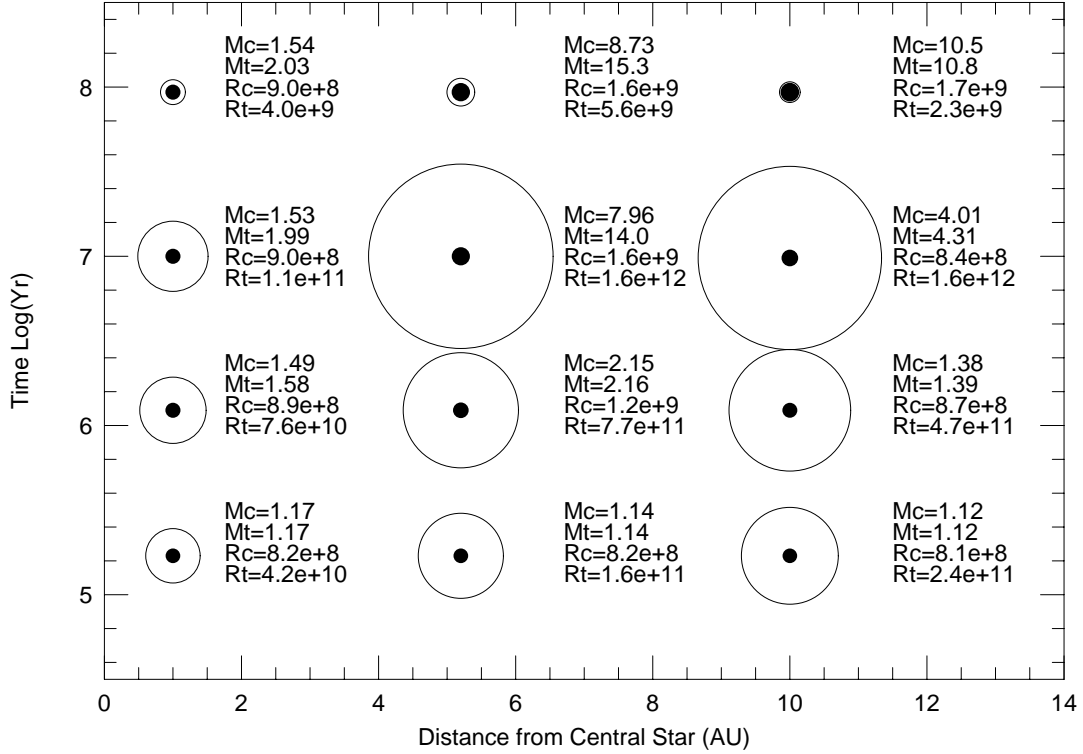}
\caption{Growth of the core and envelopes of planets forming in a 
disk surrounding a 0.4 $M_{\odot}$ star. The planetary orbits are
circular with radii $a$ = 1, 5.2, and 10 AU. For each radius, models
are shown for times $t$ = 0.17 Myr, 1.25 Myr, 10.1 Myr, and 100 Myr.
At $t$ = 100 Myr, the disk has dissipated and the planets have
undergone Kelvin-Helmholtz contraction to (nearly) reach their final
sizes. The figure lists the core mass $M_c$ and the total mass $M_t$
(in Earth masses $M_{\oplus}$); the figure also lists the core radius
$R_c$ and the total radius $R_t$ (in cm). The core sizes and total
radii are plotted with an applied scaling proportional to $r^{1/3}$.  
Protoplanets orbiting M stars should reach their maximum radius at 
time $t \approx$ 10 Myr, corresponding to the time of disk dispersal. } 
\end{figure}

\end{document}